\documentclass[11pt,twoside]{article}

\usepackage{asp2006}
\usepackage{graphicx}
\usepackage{epsf}
\usepackage{psfig}
\usepackage{lscape}

\begin{document}
\title{The DDO Close Binary Spectroscopic Program}   
\author{Slavek M. Rucinski}                           
\affil{Department of Astronomy and Astrophysics,
University of Toronto, Toronto, ON, M5S~3H4, Canada}  

\begin{abstract}                                    
The survey of radial velocity orbits for short period ($P < 1$ day),
bright ($V < 10$, with a few fainter stars) conducted at the David
Dunlap Observatory in the last 9 years before its closure in 2008
included 162 binaries and
resulted in 150 SB2 orbits and 5 SB1 spectroscopic
orbits thus becoming one of the main legacies of DDO.
The paper summarizes the main results from the survey.
\end{abstract}

\section{The historical background}

The David Dunlap Observatory, part of the Department of Astronomy and
Astrophysics, University of Toronto, served as a major Canadian astronomy
center between 1935 and 2008. In the late 1980's, a decision was made to
close the observatory and reallocate the resources into more modern
branches of astronomy. The radial velocity, short-period binary
program described here was conducted during the last
years of the DDO existence as a research facility, before its closure
on 2 July 2008.

Radial velocities (RV) of binary stars were always in the center of
research interests at DDO. The program described here
continued these interests, but its main pragmatic rationale was to
utilize the 1.9 m telescope for a useful program which could have
tangible results in a limited amount of time.
In the early 1990's, Dr.\ Hilmar Duerbeck and the author
suggested to the DDO staff a simple service
program of occasional observations of several tens
of EW (also called W~UMa or contact) binaries in
order to provide the missing radial
velocity component to combine with Hipparcos tangential velocities.
The Telescope Operator of that time was
Mr.\ Wen Lu; he went further and  obtained full orbital coverage
for several binaries. The author of this paper came back to
Toronto from Hawaii in 1999 and helped
Mr.\ Lu to analyze and publish the first batch of 20
orbits [DDO-1] and [DDO-2]\footnote{For simplicity
of referencing, papers based on DDO data
will be marked below and in the reference list by square brackets.}.
As the observatory kept on existing with a typically yearly
horizon for its closure,
the subsequent publications started to slowly acquire
a shape of a more or less systematic survey; this took place at the
time of the papers [DDO-4] or [DDO-5].
In addition to the DDO series,
a number of additional papers dealt with  special or unusual objects,
but we did try to adhere to the 10-orbit per paper format and
most orbits were published that way.

\section{Completeness of the survey}

\begin{figure}[h]
\plottwo{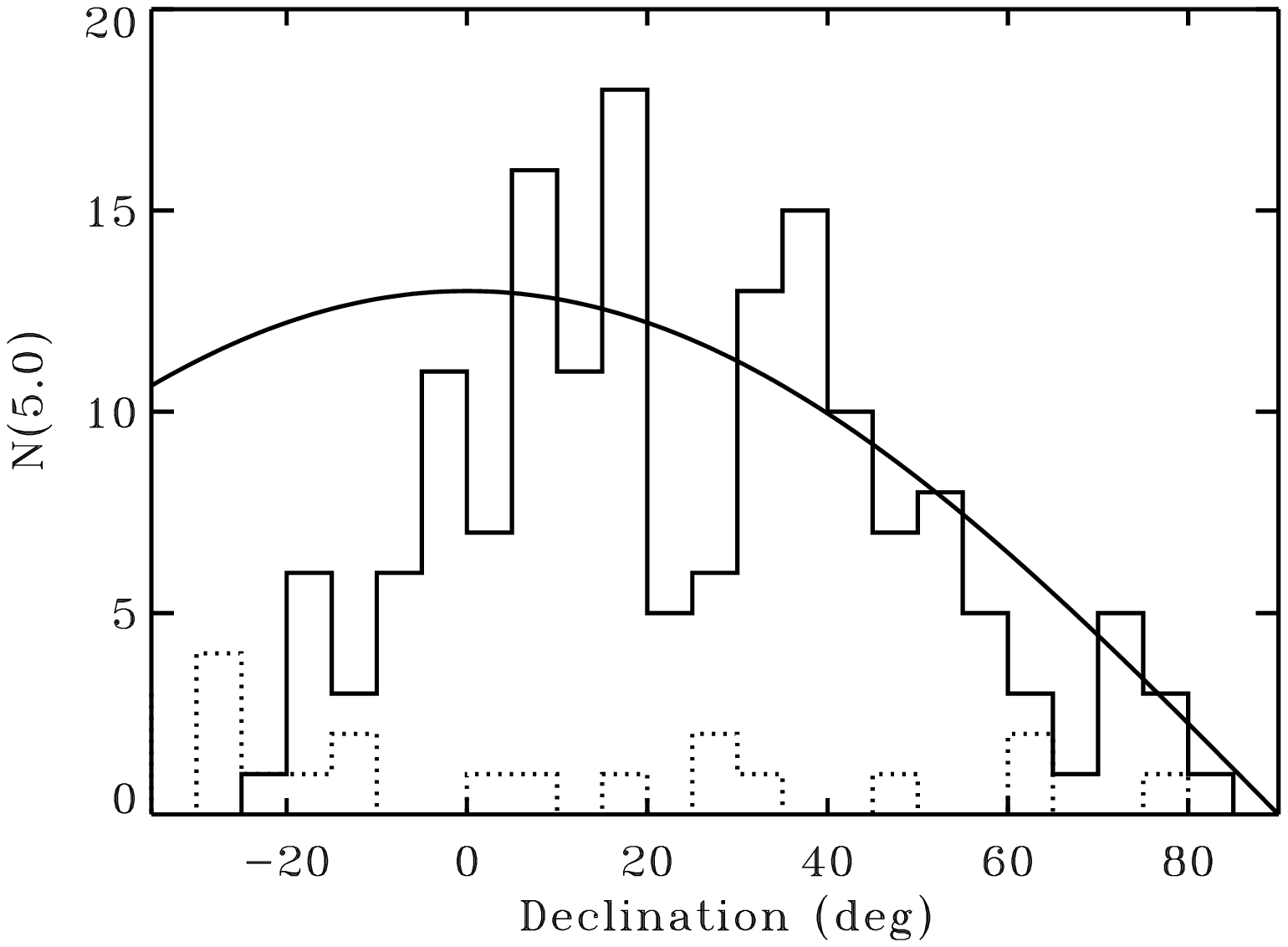}{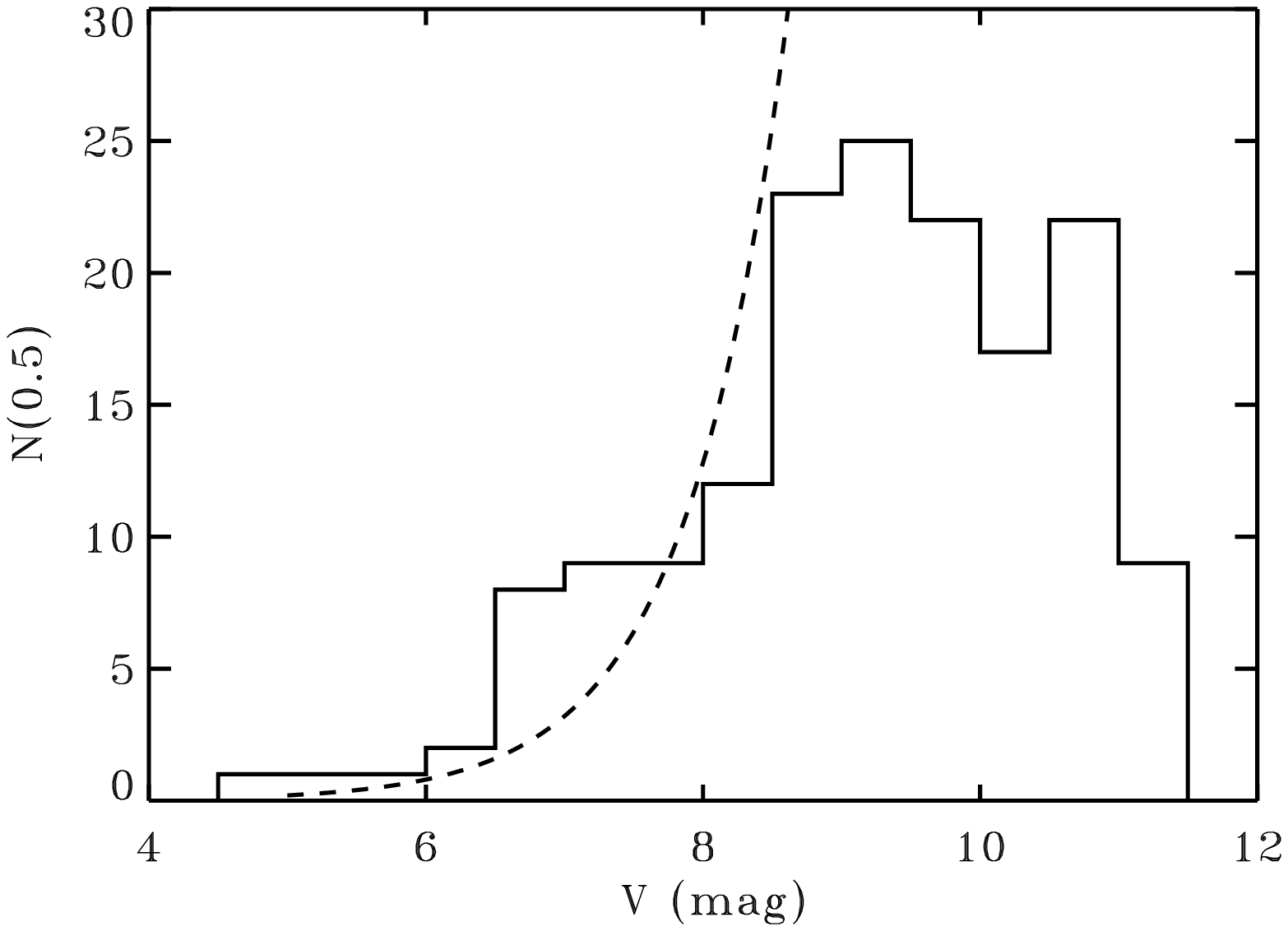}
\caption{The known survey selection effects:
The left panel shows the number of objects in 5 degree intervals of
the declination. The expected shape of the distribution is given
by the continuous cosine curve. The systems which were {\it not\/}
observed are marked along the horizontal axis by a dotted line.
The right panel gives the number of object per half-magnitude
intervals. The continuous, broken-line curve approximates
the expected $4 \times$ per magnitude increase rate.}
\label{fig1}
\end{figure}

Given the haphazard organization and the uncertain time limit of the
survey, it turned out to be surprisingly complete one exceeding the
level of 90\% completeness in terms of extant photometric
discoveries. In the end, we tried to include all known contact (EW),
semi-detached (EB) and detached (EA) binaries with periods shorter than
one day, brightness above 10th magnitude and with accessibility
limits of our telescope. Figure~\ref{fig1}
shows the number distribution of our targets versus their celestial
declination and brightness.
Obviously, targets at negative declinations are under-represented,
but this is not entirely due to difficulties of observing to the
southerly direction, directly over the bright Toronto,
but is also partly due to the smaller number of photometric
discoveries in the southern sky for stars fainter than 8th
magnitude. Some stars were simply never
observed, either because we considered the literature
data good enough or somehow did not observe them before the DDO
closure. We are aware of 14 such
objects\footnote{The 14 binaries which were not observed:
VW~Cep, BW~Dra, BV~Dra, AC~Boo, V1073~Cyg,
ER~Vul, V781~Tau, U~Peg, DW~Boo, VZ~Psc, YY~Eri, ES~Lib,
DX~Aqr, PP~Hya.}. Partial or inconclusive data for 12 binaries
are described in the last, larger publication [DDO-15] which broke the
10 stars per paper format.

As is visible in Fig.~\ref{fig1}, at the bright end,
the numbers of objects increase rapidly with magnitude, as
expected, but only to about 8 -- 9 magnitude. The abrupt break in
this rise is caused by an absence of fainter objects with
amplitudes below $\simeq 0.02 - 0.03$ mag.
The most complete photometric survey to this level of variability
is that of the Hipparcos satellite, but it is
complete only to about 8 magnitude. We note that we found many triple
systems among the Hipparcos low-amplitude (``diluted'')
photometric variables which turned out to have large RV semi-amplitudes;
we would never study them if we limited ourselves to
large photometric amplitude objects only.

\section{The observations}

Although -- in the end -- we collected some 12,000 spectra,
the yearly ``horizon'' for the DDO continuation
forced us to survey only the shortest period ($P < 1$ day) and brightest
($V < 10$) binaries visible from Toronto, a task which seemed to be
within reach in a few years and indeed was almost achieved.
The first limit was just to confine the survey scope and thus its
duration, but the second limit was driven by quality of
the spectra ($S/N >30-50$) at the required
highest resolution power of the
Cassegrain spectrograph ($R \simeq 12,000 - 15,000$) at
$V \simeq 11$, with some margin for poorer quality nights.

In the middle of the program,  the old, scratched grating
of 1800 lines/mm was replaced by a new 2160 line/mm grating; this did not
increase the resolving power (set by the spectrograph optics),
but improved the sampling. We also changed
the main CCD detector of 1024 pixels at 19 $\mu$m by a longer detector
od 2048 pixels at 13.5 $\mu$m.

\begin{figure}[h]
\begin{center}
\includegraphics[width=75mm]{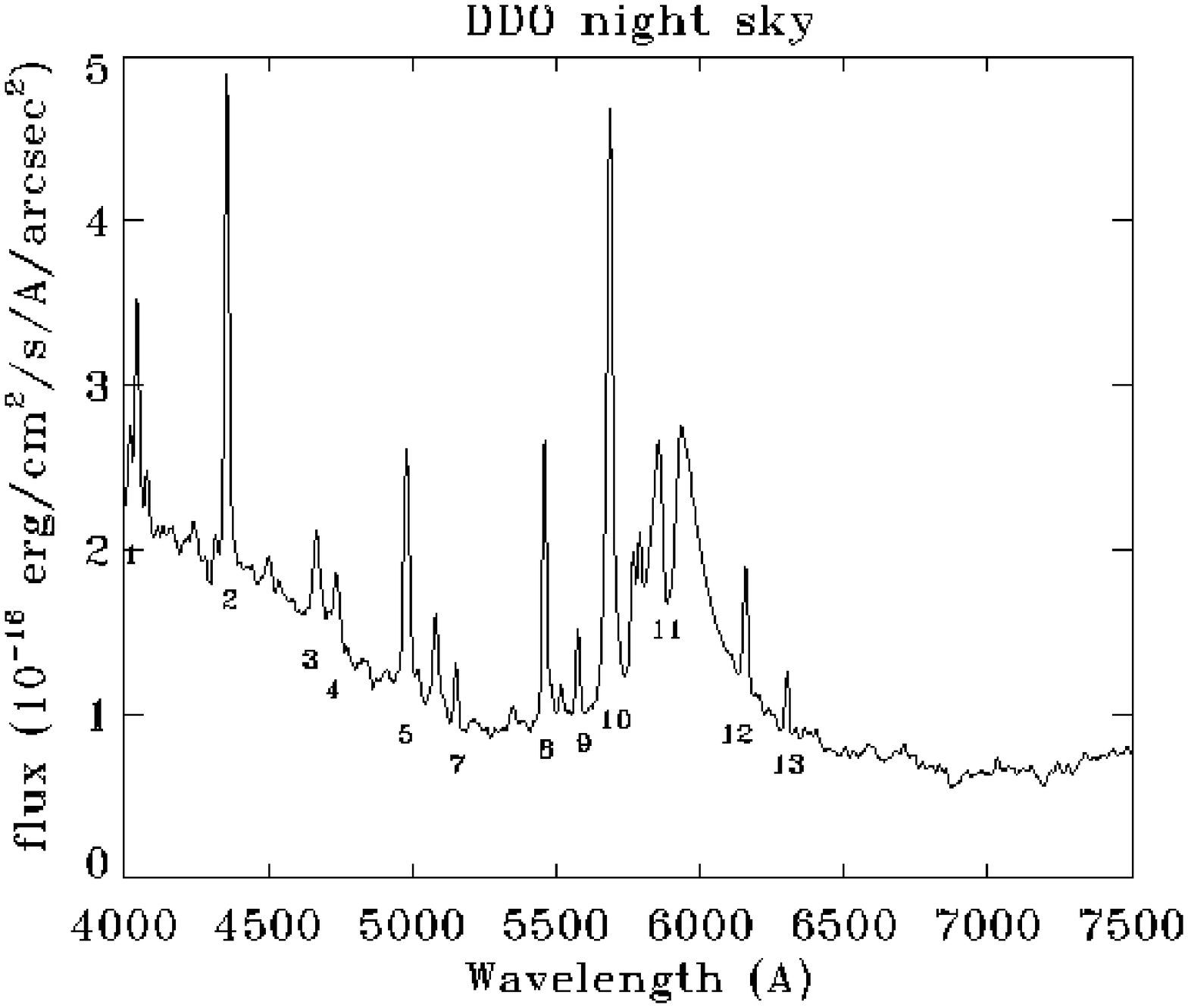}
\caption{The Toronto sky is very bright. This calibrated, low
resolution spectrum was obtained by Dr.\ R. M. Blake.
It is dominated by the strong high-pressure lamp sodium feature.
The Mg~I 5184 \AA\ triplet used in this program
is situated in the brightness minimum,
close to the feature marked as \#7.
This particular spectrum shows a weak lunar
light contamination in the blue.
}
\label{fig2}
\end{center}
\end{figure}

Most of the survey was done in the Mg~I triplet region at 5184 \AA, within
a spectral window about 220 \AA\ wide. This region is not only rich in
spectral line in late-type spectra but its location in a city sky minimum
gave another advantage in terms of the quality of spectra (Fig.~\ref{fig2}).

\section{The broadening functions (BFs)}

The broadening functions (BF) technique was developed
and improved during the survey as we attempted to analyze progressively
more complex spectra. Its first use dates back
to an analysis of AW~UMa using CFHT spectra \citep{AW1}. Later,
a short description \citep{Victoria99} stimulated further
development by others,  a technique utilizing model spectra
known as the Least Squares Deconvolution (LSD).
These techniques are not identical: The BF technique uses
standard star spectra while the LSD technique uses model spectra.
While the LSD  gives smoother deconvolution results,
we valued the ease of tying our RV's to the standard stellar
velocities system as well as a simple reference to the spectral
sequence: Integrated BF intensities at the Mg~I triplet
change monotonically with the spectral type and can be used
as an independent check on the type of the star (in addition to
the color and the Sp Type estimate). This interesting aspect
will be discussed in a separate investigation.


\begin{figure}[h]
\plottwo{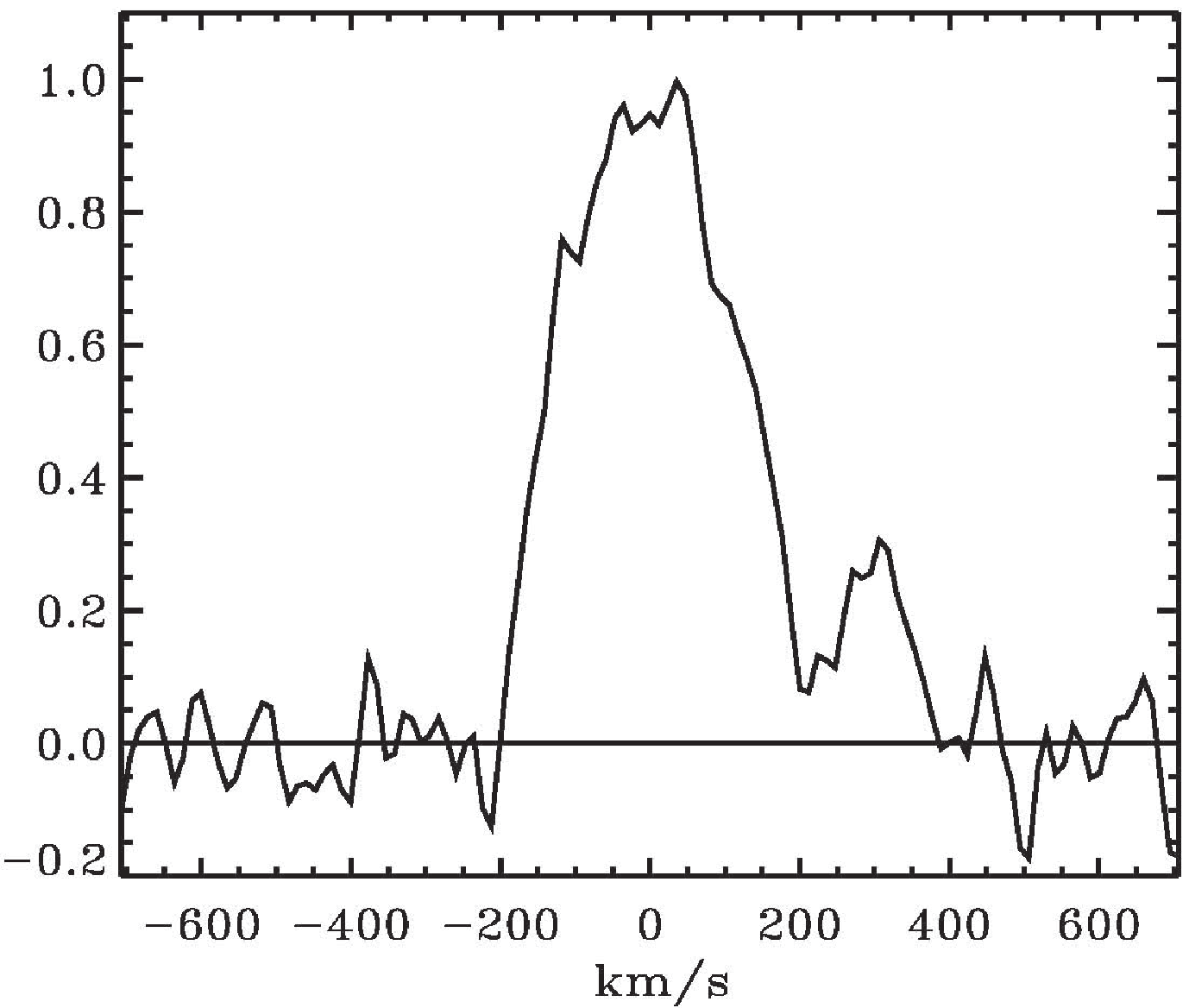}{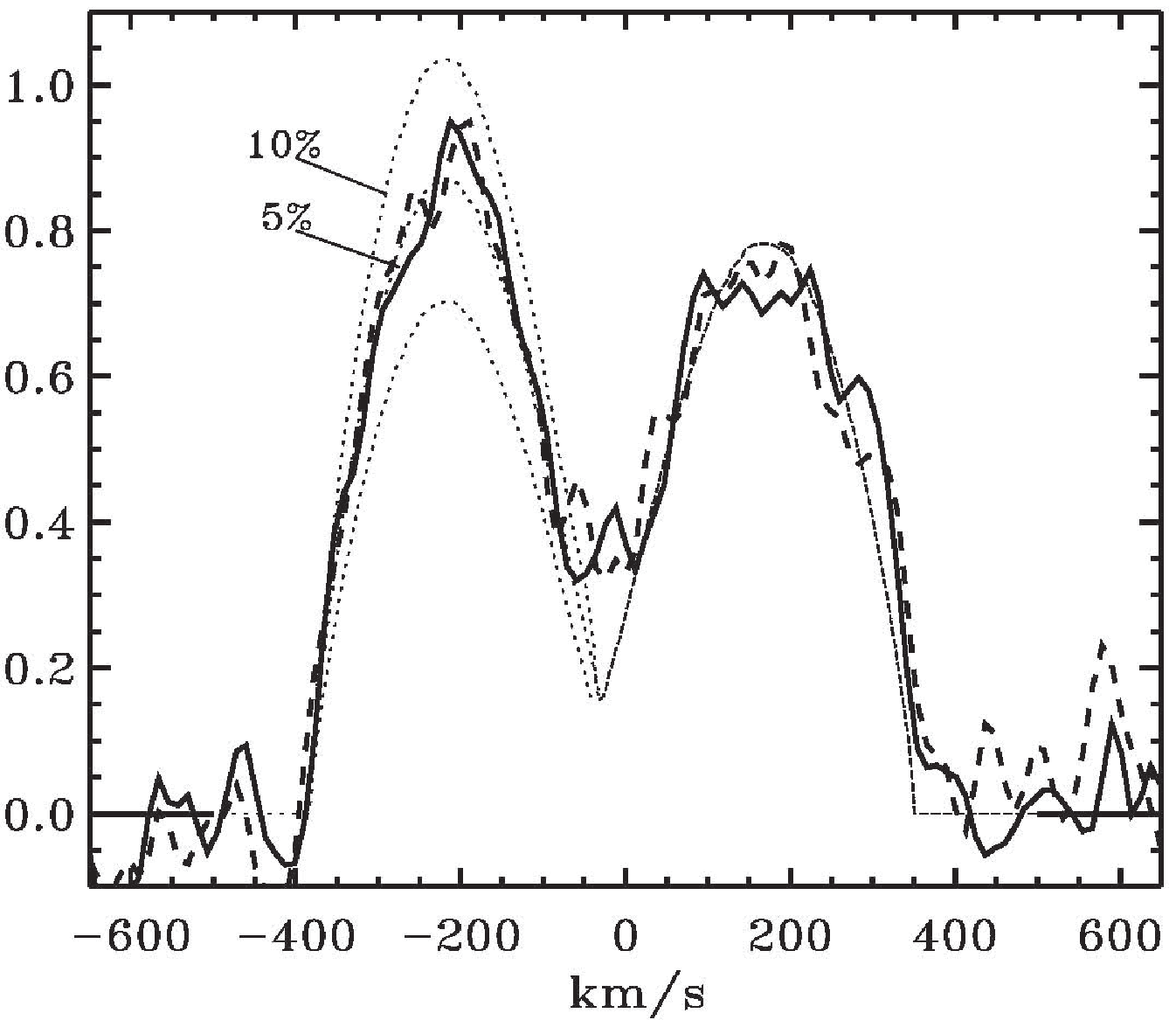}
\caption{The broadening functions. The left panel:
The BF for the extreme mass-ratio system SX~Crv
at its orbital quadrature. The value of $q \simeq 0.06$ is
uncertain because of difficulties with measuring the minuscule
RV shifts of the {\it primary\/} component.
The right panel: The BF for SW~Lac where
we directly see the secondary-component surface-brightness
enhancement (the taller, narrower peak),
here parameterized by the relative temperature increase,
$\Delta T/T$. The dash-dotted line without a label
is for the unmodified ``Lucy model'' with $\Delta T/T =0$.
}
\label{fig3}
\end{figure}

The idea behind the BF is simple: With a template (standard star)
spectrum representing all natural broadening effects, to de-convolve
the binary spectrum for all effects
caused by radial velocity effects of rotational broadening and
orbital motion. The convolution of the template spectrum
$T(n)$ with $B(m)$ ($n/m > 1$, preferably several times)
into the broadened spectrum $P(n)$, is written as
a set of linear equations and then solved using least squares.
The convolution integral transform is written as an array operation:
\begin{equation}
P(\lambda ') = \int B(\lambda ' - \lambda) \: T(\lambda) \: d \lambda
\,\,\,\,\, \Rightarrow \,\,\,\,\,
\vec{P} = \mathbf{D} \, \vec{B}
\end{equation}
where the rectangular array $\mathbf{D}$ contains
the appropriately shifted vector $\vec{T}$ as its columns.
The broadening function is represented by a vector
of the unknowns, $\vec{B}$; the array $\mathbf{D}$
has dimensions $m$ by $n-m+1$.
Thus, the system of over-determined linear equations for
$\vec{B}$ is:
\begin{equation}
P_i=\sum_{j=0}^{m-1} T_{i+m-j} \, B_j \qquad\mbox{with}\qquad
i=m', \ldots, n-m'-1
\end{equation}

The result, the Broadening Function,
looks and functions (Figure~\ref{fig3})
very much like the cross-correlation function
(CCF) but is in many ways superior to it: (1)~it has better
resolution, (2)~it is linear, so that it is a simple mapping
of stars into the RV domain, (3)~the respective peaks can be
integrated and do not require any calibration
for the $L_2/L_1$ ratio (as does the CCF route), (4)~does not show
any fringing of the baseline which is characteristic for
the CCF.

The BF technique is not as easy to use as the CCF though:
The proper length of the BF window (not too long and not too short)
should be adjusted beforehand by running a preliminary CCF; also,
very long spectra are difficult to use in the array solution
(with echelle spectra, it is better
to work order by order). Details of the BF implementation
and IDL scripts are available in author's Web
page\footnote{www.astro.utoronto.ca/$\sim$rucinski}.


\begin{figure}[h]
\plottwo{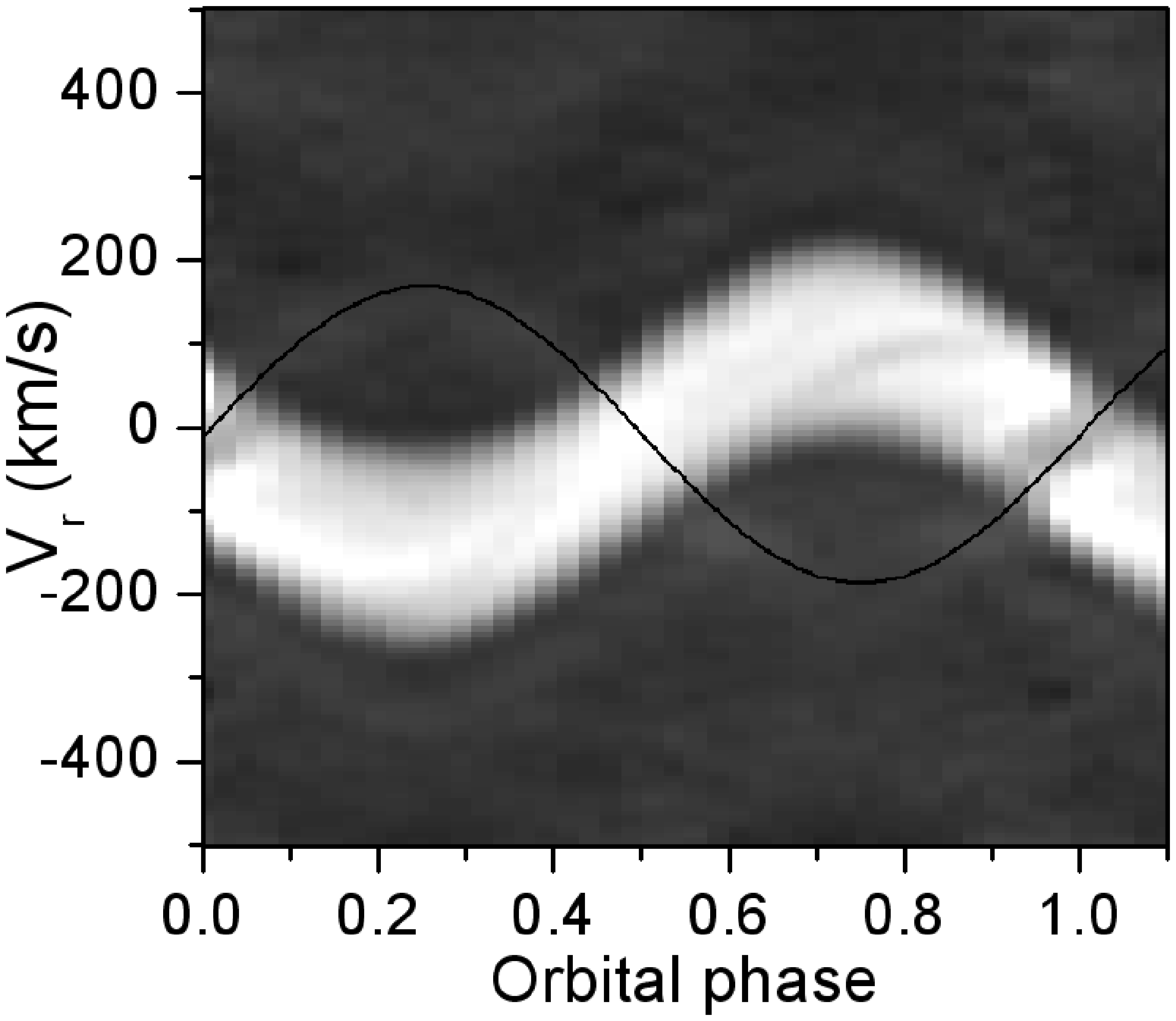}{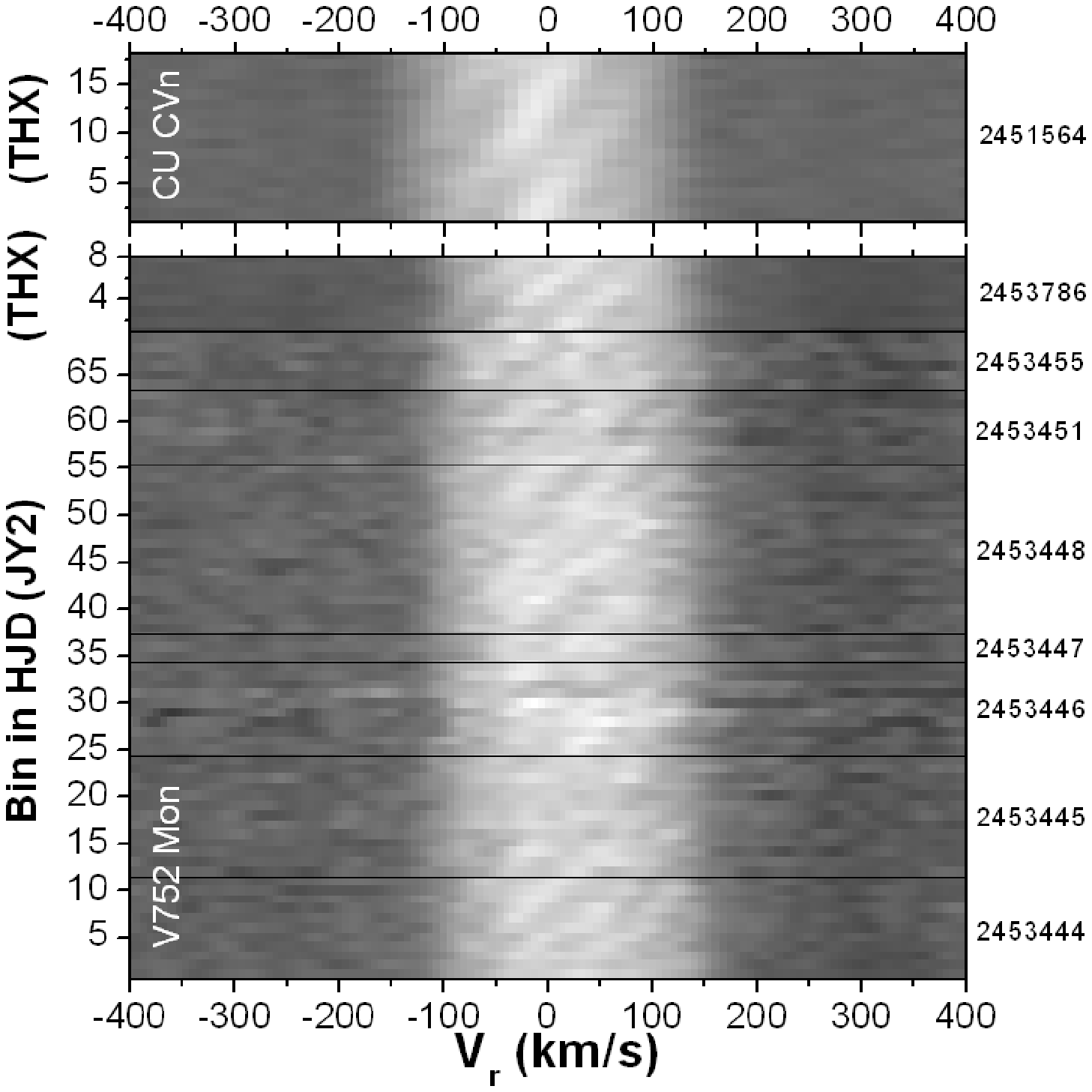}
\caption{The temporal or orbital phase
correlations in broadening functions
are best visible when they are arranged into 2-D images.
Left panel: The secondary's transit and 
photospheric spots on the primary of the detached binary XY~UMa
[DDO-12].
Right panel: Non-radial pulsations of $l \simeq 6-8$ manifest
themselves on an apparently single stars (dominating primaries or
third stars) in CU~CVn and V752~Mon as slanted ripples [DDO-15].
}
\label{fig4}
\end{figure}

\section{Applications of BFs}

The BF technique was certainly crucial in our ability to
study so many double-line binaries: The number ratio of 150 SB2's
versus 5 SB1's says it the best; the few SB1's are
genuinely single-line binaries, most likely contain white or brown
dwarfs or massive planets. Thanks to it, we have been able to
detect several contact systems with very small mass ratio,
$q = M_2/M_1$ at the level of 0.1 and below. Among them
the system SX~Crv [DDO-5] which appears to have the mass
ratio even smaller than that of AW~UMa, perhaps as small
as $q = 0.06$ (Figure~\ref{fig3}).
Paradoxically, for such extreme systems,
the difficulty is not in detection and velocity measurements
of the low-mass component, but in determination of
velocities from the large, wide lobe of the heavily
broadened primary feature which does not move much.
The only proper approach would
be to model such BFs, to relate velocity centroid
determinations to the velocity of the primary mass center.

Modeling of BFs has a potential of studying deviations
from the assumed binary model. One such an application would be
to look into the still unexplained cause of the W-type light
curve deviations from the Lucy model, so well visible
in the BFs of SW~Lac [DDO-10] and apparently reproducible
by a simple increase of the surface temperature by a few
percent (Figure~\ref{fig3}). Also, the BFs,
when arranged in time or orbital phase can show
features which are hard to detect in other ways,
particularly weak, drifting photospheric
dark spots and non-radial pulsations (Figure~\ref{fig4}).


\begin{figure}[h]
\plottwo{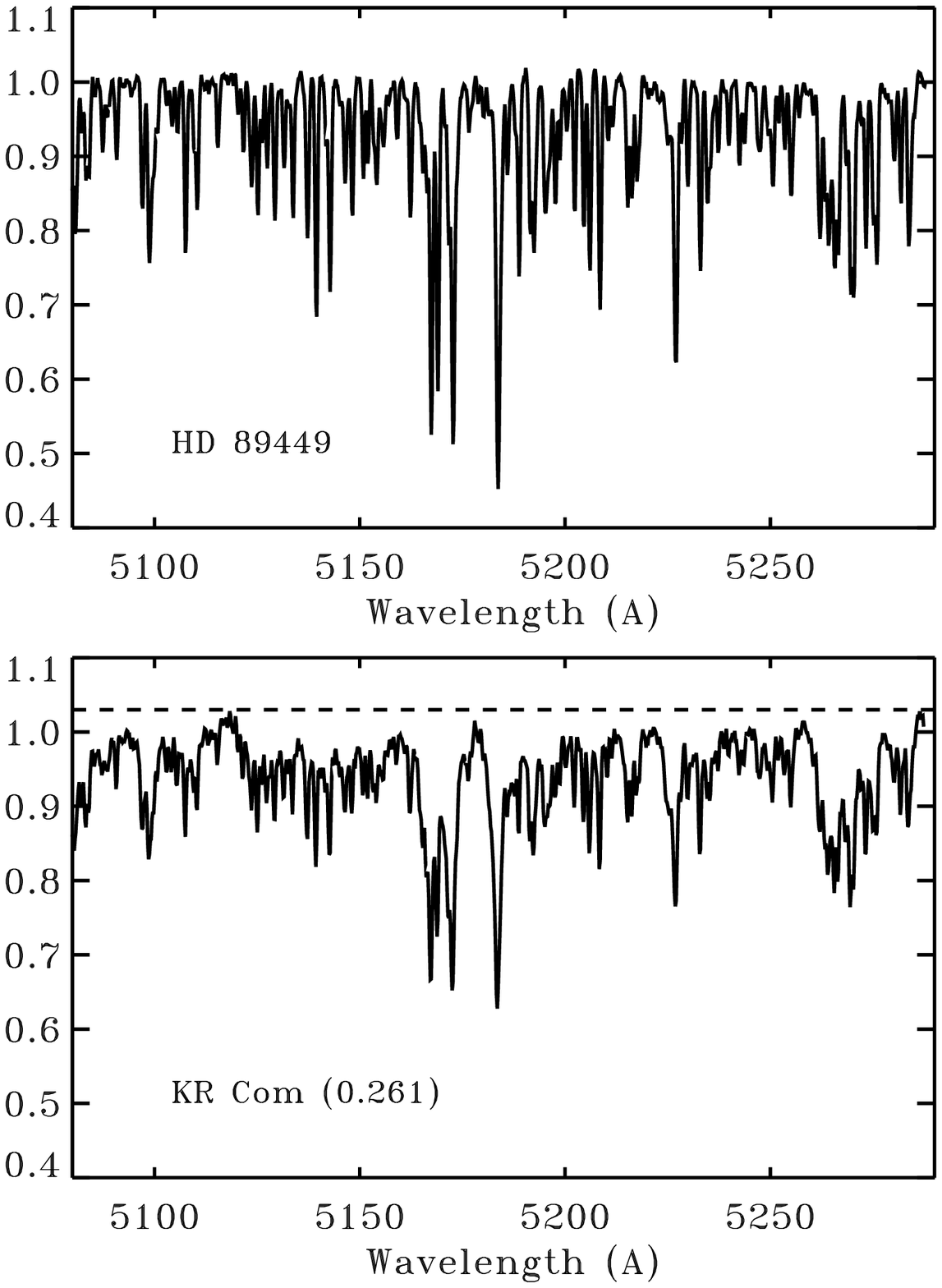}{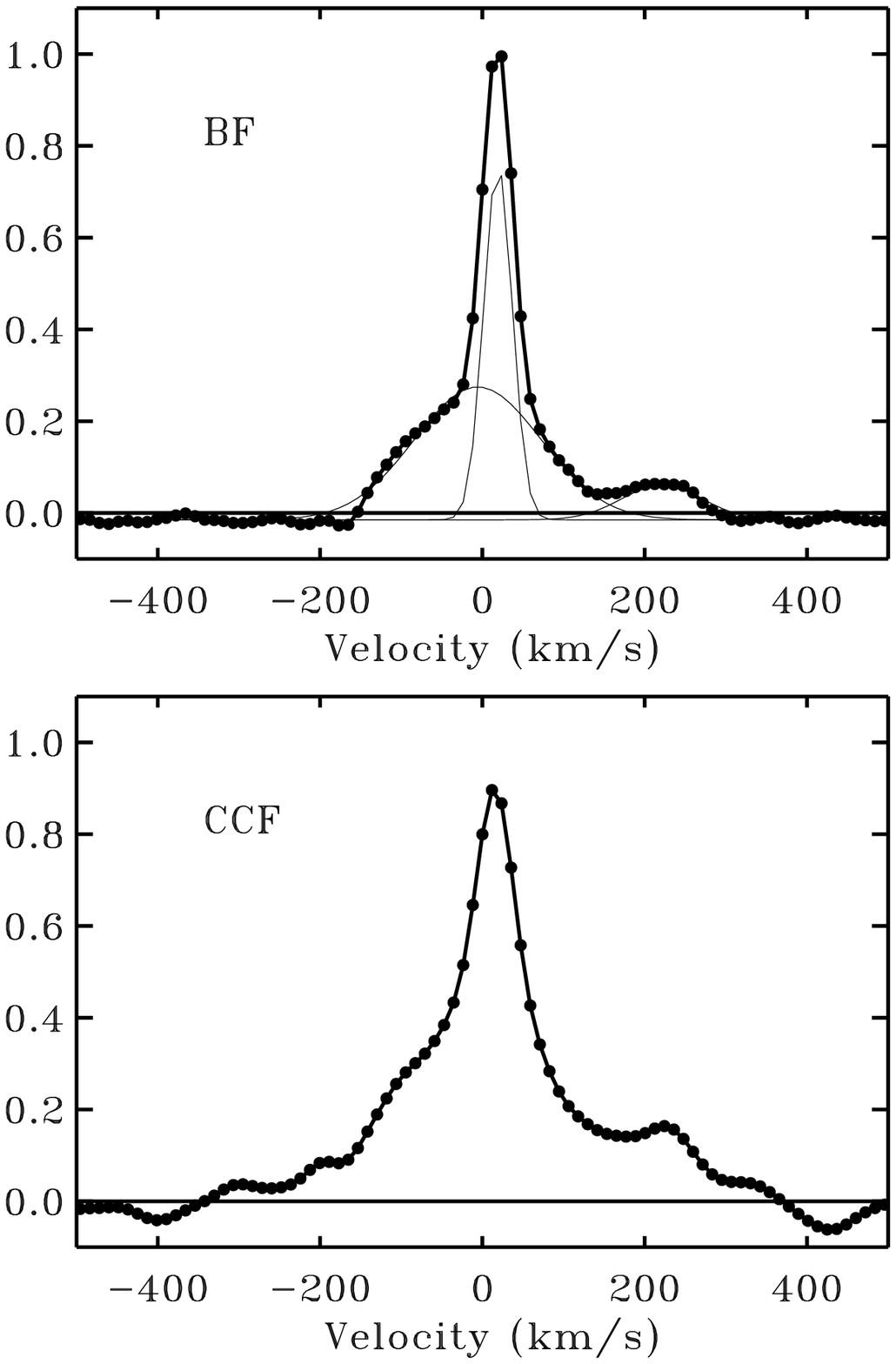}
\caption{A typical triple system, KR~Com. The left panel shows
spectra of the template and of KR~Com at the orbital phase 0.261.
The star was previously considered chemically peculiar because
of its strange continuum, now explainable by heavy blending
of the close binary lines. The
BF and CCF functions are compared in the right panel. Note, that
the BF baseline is slightly negative. This is due to the necessity to
use the pseudo-continuum for spectral normalization/rectification.
This does not affect the intensities because the BF is linear,
but does affect estimates of $\beta = L_3/L_{12}$: The slowly rotating
component and the binary have
spectral continua whose hight ratio does
not correspond to the luminosity ratio; thus, usually,
$\beta_{obs} > \beta_{true}$. For the KR~Com spectra shown here,
$\beta_{obs} = 0.56$ so that the third star has a comparable
brightness to components of the binary, but certainly dominates in the
appearance of the spectrum.
}
\label{fig5}
\end{figure}

\section{Triple systems}

At first, we had so many SB2 spectra that we did not analyze
any triple and multiple systems, but later we
realized that the BF approach is particularly useful for situations when
more spectral features are visible together because it offers
very high fidelity of the information extraction. Also, with
the Hipparcos low photometric amplitude binary detections,
we were confronted with the increasing number of systems
with relative large $K_i$ semi-amplitudes but showing
two or more BF additional peaks of stationary or semi-stationary
components.

Realization that triple systems are so common in our survey
led to resurrection of the old idea \citep{RK1982}
that close binaries exist {\it because\/} they are
in triple/multiple systems. This has led to a survey of
literature data [Triples-1] and to an adaptive-optics search
for close companions
[Triples-3]. In the present context, the most interesting of
the DDO data utilization was the project involving averaging
of several of co-aligned spectra to detect stationary RV components
of companions [Triples-2].
Because of the limited space,
we refer the reader to the papers of this series for details.
We only note that we have indications that binary systems
with periods shorter than one day show the apparent frequency
of companions $>60$\% which is consistent with 100\% incidence.

\section{The mass ratios}

Among 150 SB2 orbits of the DDO program,
121 were of contact (EW) systems;
the rest were semi-detached (EB; 14), detached (EA; 10) and
uncertain in type, ellipsoidal variables (Ell, 5).
The semi-amplitudes $K_1$ and $K_2$ for the SB2 binaries,
although probably the best currently available,
may be  affected by the way we measured them. Both,
the Gaussian (used up to [DDO-10]) and rotational profiles
(used from [DDO-11]) are symmetric functions, while peaks in the
BFs are certainly not symmetric. The only proper approach would
be to model the BFs in full, assuming a contact,
semi-detached or detached model. But, for many systems, the
photometric data were not available;
besides, we simply had no resources for complex,
combined radial-velocity and photometric solutions.

\begin{figure}[h]
\begin{center}
\includegraphics[width=80mm]{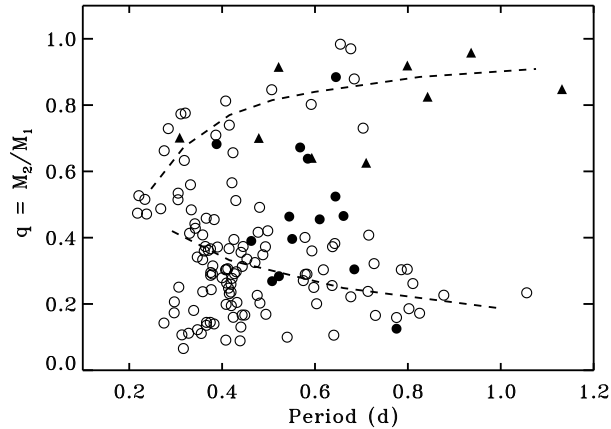}
\caption{The spectroscopic mass ratio versus the orbital
period for all SB2 systems of the survey. The symbols are:
open circles for EW, filled circles for EB and triangles
for EA systems. The high frequency of contact systems is
well known for orbital periods around $P \simeq 0.3 - 0.4$,
but note the absence of binaries for $P > 0.6$ days
and moderate mass ratios. The two sequences,
as marked approximately by the broken lines, correspond to
preferred mass ratios of
EW + a few EB (lower) and EA + a few EB (upper) types.}
\label{fig6}
\end{center}
\end{figure}

The mass ratios, determined  from $q = K_1/K_2$
seem to be least affected by the way we determined the
radial velocities. A plot of mass ratios versus the
orbital period (Figure~\ref{fig6})
shows an unexpected avoidance of moderate values of
the mass ratio (roughly $0.3 < q < 0.8$) for periods
longer that 0.6 day. One can even see  two sequences,
(1)~of detached (EA) and perhaps semi-detached
(EB) systems with large mass ratios and (2)~of contact systems
with mass ratios approaching smallest detectable values.
If one were brave, one can notice a convergence of both
sequences at the very short-period end where
they tend to the same mass
ratio of $q \simeq 0.5$. Of course, the number of
binaries remains small so that a statistical fluke is not
excluded. But, if confirmed, the relation shown
in Figure~\ref{fig6} is a new, unexpected and important result.

\begin{figure}[h]
\plottwo{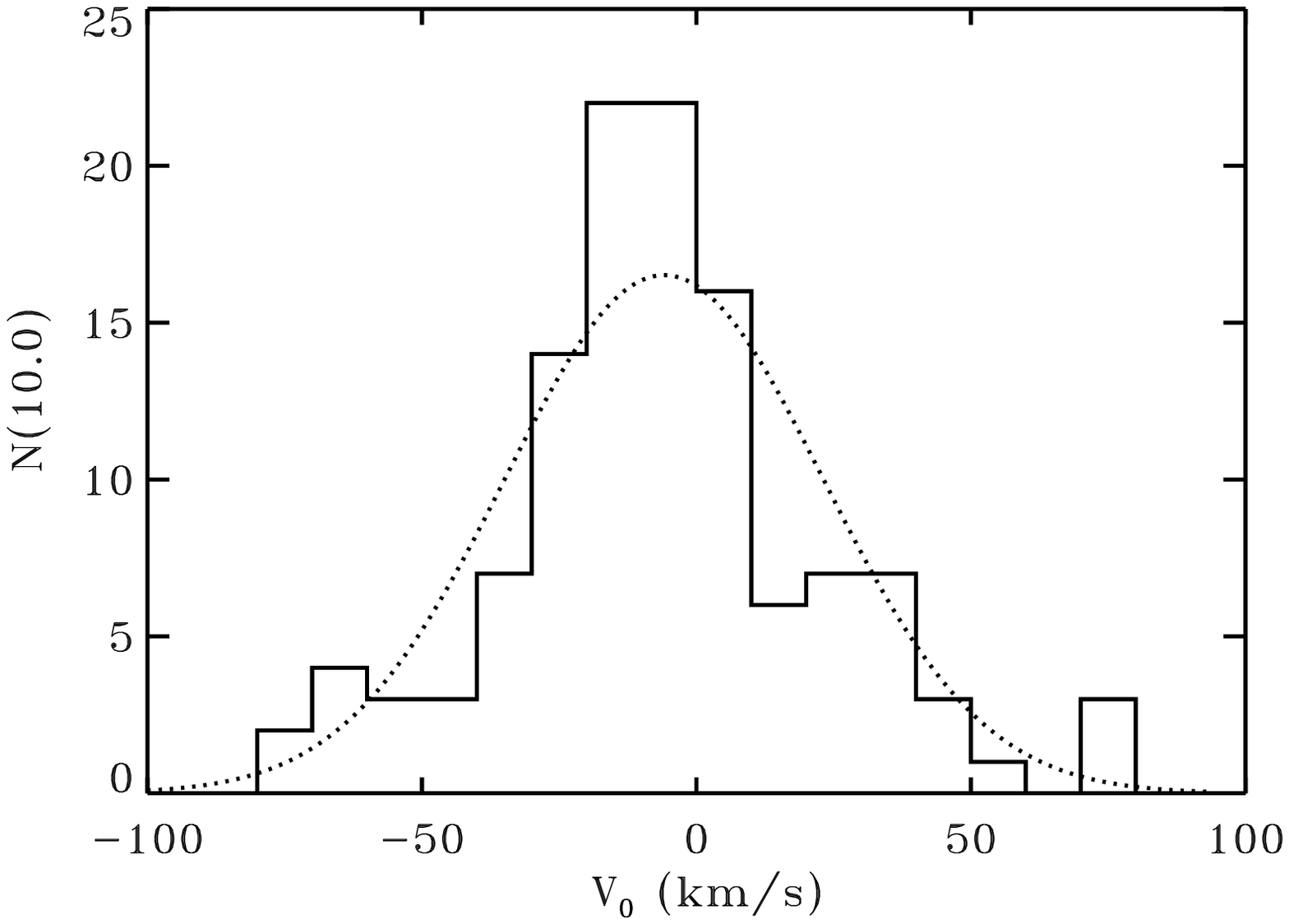}{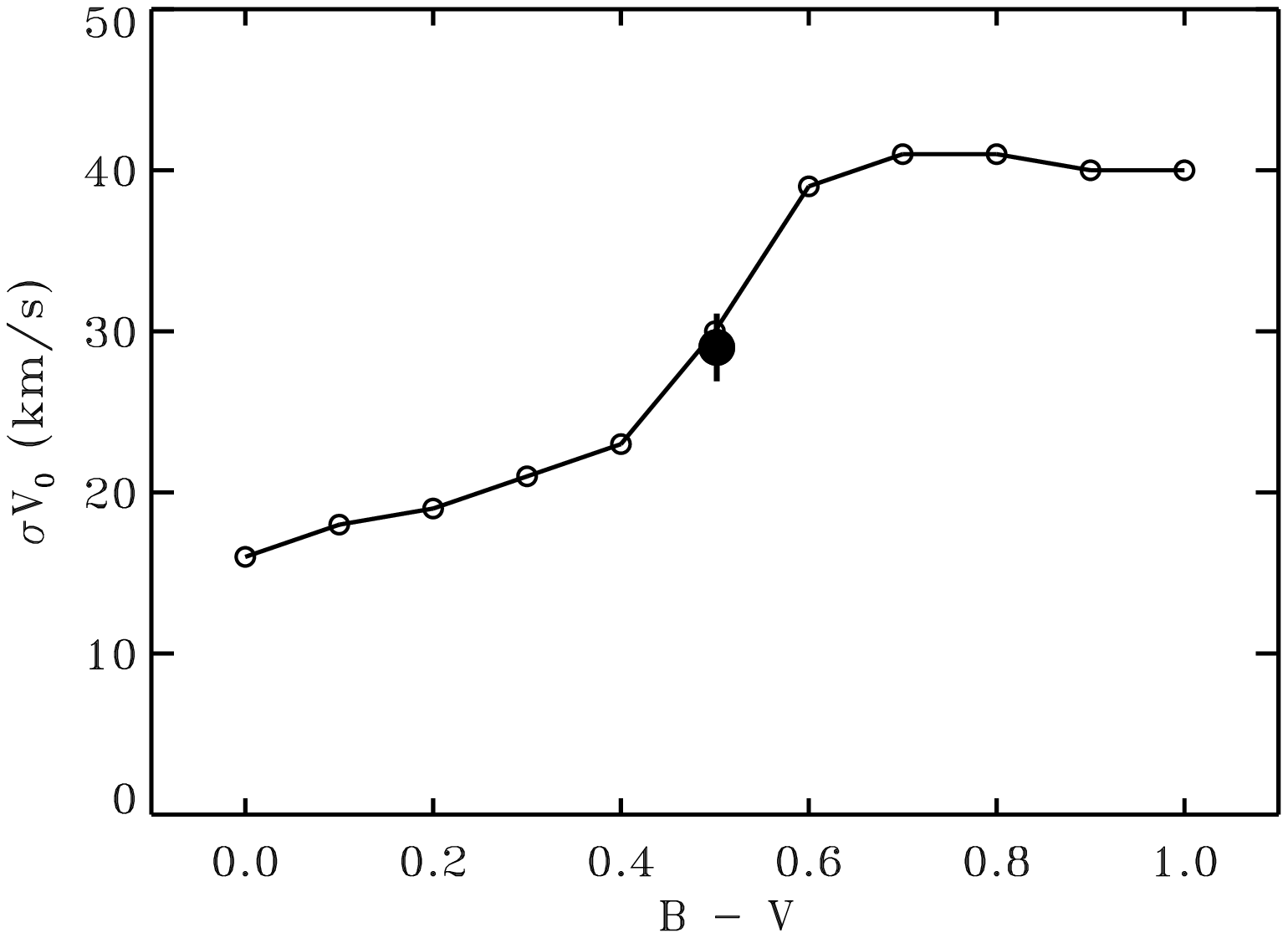}
\caption{The distribution of the $V_0$ (center of mass)
velocities of contact binaries is shown in the left panel.
The right panel gives the field, Main Sequence one-dimensional
velocity dispersion versus the $B-V$ color index based
on the \citet{DB1998} study. The dispersion
for EW systems is shown for their mean $B-V$ as a filled circle;
the numbers of EW systems per color bin are small but
the MS progression appears to be present there. This result indicates
that W~UMa binaries are not substantially older than most of
the field (old disk) stars.
}
\label{fig7}
\end{figure}

\section{Center of mass velocities}

Are contact systems old? Many lines of evidence suggested
that it takes long time for initially close, but detached
binaries to come into contact and to live a new ``contact'' life.
Previous kinematic investigations by
\citet{GB1988} and \citet{bilir2005}\footnote{DDO results
up to and including [DDO-9] were used in the \citet{bilir2005}
study, in addition to data from other, generally poorer
sources. Such studies can now benefit from subsequently
published DDO data.}
indicated that center-of-mass velocities of EW
binaries have a large dispersion, a property which could be
interpreted as due to the advanced age.

The new statistics based on all
121 contact systems of the survey does not
confirm any excess in the velocity dispersion.
The velocity dispersion is
$\sigma V_0 = 29.0 \pm 2.1$ km~s$^{-1}$ which is {\it not
different\/} from the dispersion of Main Sequence stars in
the field, as based on the data given in \citet{DB1998}.
We cannot exclude a possibility that the picture is more
complex with two contributing distributions of different widths, 
as indicated by the actual shape of the $V_0$ 
histogram in Figure~\ref{fig7}.

\begin{figure}[h]
\begin{center}
\includegraphics[angle=270,width=100mm]{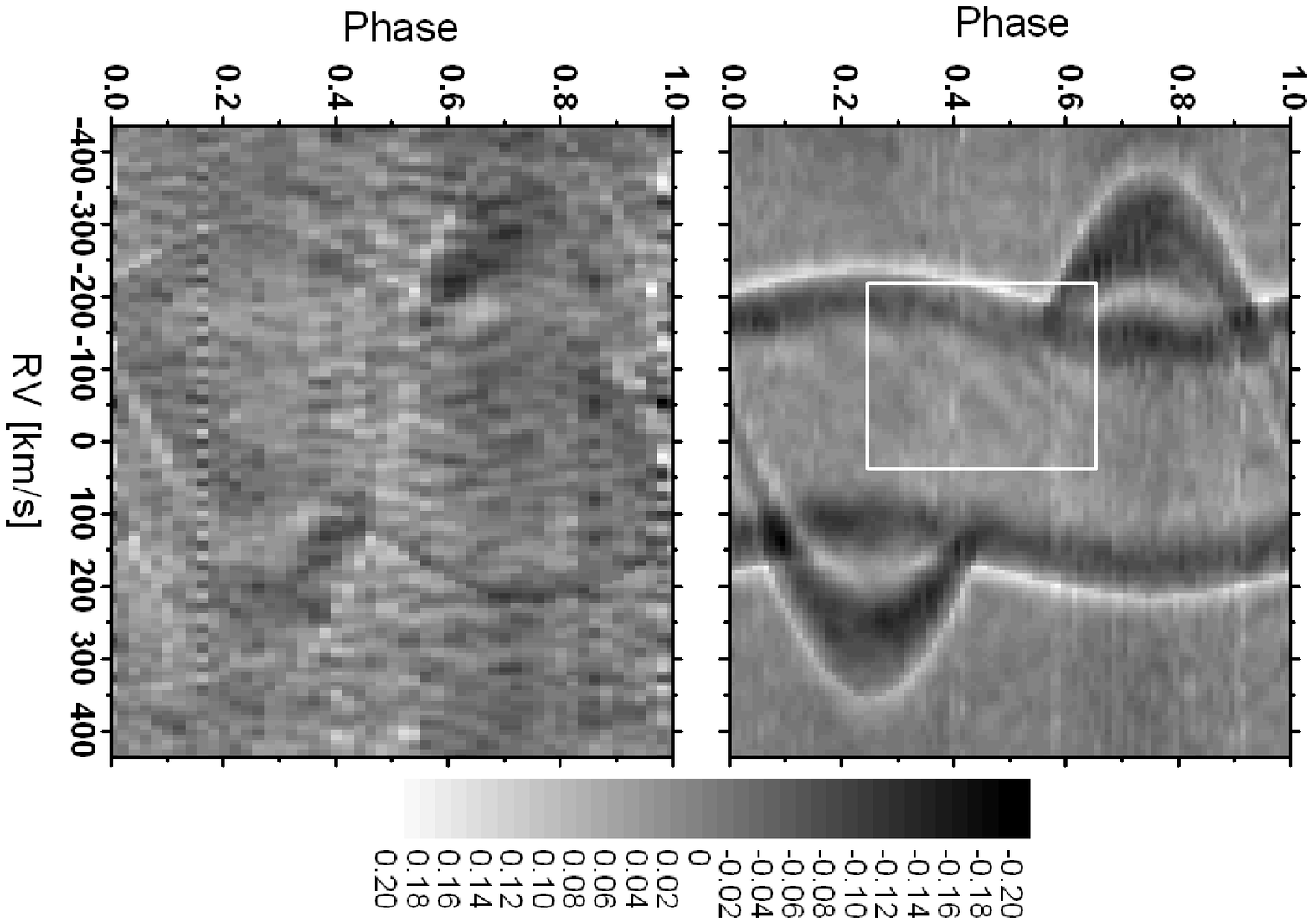}
\caption{The grey scale figure
shows deviations from the model BFs arranged in
orbital phase (horizontal) for V566~Oph (left) and
AW~UMa (right); the vertical scale gives velocities.
The box marks the region where spots or non-radial
pulsations on the AW~UMa primary are visible as inclined
ripples.
}
\label{fig8}
\end{center}
\end{figure}

\section{The AW~UMa system; is the contact model right?}

The AW~UMa binary, known also as Paczynski's
star\footnote{\citet{BeP1964} discovered AW~UMa
in 1964  and immediately
recognized its importance; this star accompanied him until
his last publication \citep{BeP2007}.} has been crucially
important for our understanding of contact binaries.
In spite of the extreme mass ratio ($q_{ph} \simeq 0.07 - 0.08$),
its light curve beautifully agrees with the
``Lucy model'' of a contact binary which fills the Roche
common equipotential and is subject to standard limb
and gravity darkening description. \citet{Wilson2008}
singled out this binary as one which can serve as an
excellent indicator of the distance, comparable
in quality to best trigonometric parallax
determinations\footnote{This view has been confirmed
in a private conversation with Dr.\ Wilson during
this conference.}.

To our surprise, the DDO detailed spectroscopic analysis
[AW~UMa] revealed large deviations from the contact model
(Figure~\ref{fig8}): (1)~As seen in the BFs, the spectral
lines are too narrow, but they do have broad
bases; (2)~The spectroscopic mass ratio estimated from
centroid motion and from modeling
cannot be reconciled with the photometric one and
it cannot be smaller than $q_{sp} \simeq 0.10$; (3)~The
secondary component appears small and asymmetric
when projected against the primary; large changes in its
shape are apparent when the orbital quadratures are compared.
The latter variability is particularly large at phases around 0.64.
It should be noted that the contact binary V566~Oph with
$q_{sp} = 0.26$ does not show any systematic deviations from
the contact model (but except for an unexpectedly
large scatter in the secondary BF, also at phases of about 0.64).

These results are puzzling and hard to explain. Is the
domain of contact binaries populated by objects with
various adherence to the contact model with some semi-detached
systems ``pretending'' to be contact binaries.
What controls the
discrepancies? As a caution: We should remember that the BFs
give us information about velocities only; these should not
be mistaken for spatial positions as these must involve models.

\section{Conclusions}

The DDO radial velocity survey of close binaries
is the most complete among such surveys with
more than 90\% of the currently photometrically recognized
close binaries with $P < 1$ day and $<10$ mag.
The survey will not be continued; the DDO has been
closed. The high success in determination of
many SB2 orbits and in detection of many companions
is partly due to the Broadening Function (BF) approach
which is a much better analysis tool than the Cross Correlation
Function (CCF). Among the new results, the unexpected
bifurcation of the $P$ versus $q$ relation requires confirmation.
The case of the AW~UMa
binary is probably the most interesting and intriguing:
The light curve beautifully agrees with the contact
model yet -- spectroscopically -- we see very serious
deviations from it.

Many further studies will utilize the DDO data.
Here, a plea to those who will use them: Please
do not do "photometric improvements" to our mass-ratios.
There is plenty of room for other investigations,
but tiny changes to the mass ratio, based on light-curve
fits are the least important and meaningful of
all what can be done... The case of AW~UMa should be taken
as a warning on how incomplete may be the picture based
on light curves alone.

\acknowledgements
The David Dunlap Observatory binary-star
survey would not be possible without the hard work and dedication of many
individuals. I would like to express my sincere thanks to
Mel Blake, George Conidis, Bryce Croll, Chris Capobianco,
Heide DeBond, Cosmas Gazeas, Toomas Karmo, Krzysztof Kaminski,
Piotr Ligeza, Wenxian Lu, Stefan Mochnacki,
Waldemar Ogloza, Bogumil Pilecki, Theodor Pribulla,
Wojtek Pych, Archie de Ridder, Matt Rock,
Piotr Rogoziecki, Michal Siwak, Greg Stachowski, Jim Thomson.

\end{document}